\documentclass[%
twocolumn,
superscriptaddress,
 bibnotes,
 amsmath,amssymb,
 prf,
longbibliography
]{revtex4}

\usepackage{graphicx}
\usepackage{color}
\usepackage{dcolumn}
\usepackage{bm}
\usepackage{hyperref}
\usepackage{bigints}
\usepackage[dvipsnames]{xcolor}

\usepackage{textcomp}

\usepackage{rotating}

\usepackage{xcolor}

\usepackage{newfloat}

\DeclareFloatingEnvironment[
  fileext=lof,
  listname={List of Extended Data Figures},
  name=Extended Data Fig.,
  placement=htbp,
  within=none,
]{extendedfigure}

\DeclareFloatingEnvironment[
  fileext=lof,
  listname={List of Supplementary Information Figures},
  name=Supplementary Information Fig.,
  placement=htbp,
  within=none,
]{SIfigure}

\usepackage{tikz}

\begin{document}

\title{Strain-stiffening elastomers fail from the edge}

\author{Nan~Xue}
\affiliation{Department of Materials, ETH Z\"urich, 8093 Z\"urich, Switzerland.}
\author{Rong~Long}
\affiliation{Department of Mechanical Engineering, University of Colorado Boulder, Boulder, CO 80309, USA.}
\author{Eric~R.~Dufresne}
\email{eric.dufresne@mat.ethz.ch}
\affiliation{Department of Materials, ETH Z\"urich, 8093 Z\"urich, Switzerland.}
\author{Robert~W.~Style}
\email{robert.style@mat.ethz.ch}
\affiliation{Department of Materials, ETH Z\"urich, 8093 Z\"urich, Switzerland.}

\date{\today}

\begin{abstract}
The accurate measurement of fracture resistance in elastomers is essential for predicting the mechanical limits of soft devices.
Usually, this is achieved by performing tearing or peeling experiments on thin-sheet samples.
Here, we show that these tests can be surprisingly thickness-dependent, with thicker samples being significantly stronger than thinner ones.
Even for a simple geometry, direct imaging of the fracture surface shows that the fracture process actually involves three distinct cracks: an inner crack, and two edge cracks.
Ultimately, samples fail when two edge cracks meet at the sample's mid-plane.
The opening angle of edge crack, $2 \theta$, determines how far the sample has to be stretched before the edge cracks meet.
Conveniently, $\theta$ is a material property that can be inferred from the elastomer's non-linear elastic response.
To yield thickness-independent fracture-test results,  sample thickness should be much smaller than the smallest lateral sample dimension divided by $\tan \theta$.
Our results have direct implications for characterizing, understanding, and modelling fracture in soft elastomers.
\end{abstract}

\maketitle


Soft elastomers can undergo large and reversible deformations \cite{creton2016fracture, long2021fracture}, making them useful in many fields ranging from soft robotics \cite{whitesides2018soft, cianchetti2018biomedical} to stretchable electronics \cite{rogers2010materials, yang2018hydrogel}.
A key consideration when using such materials is how they fail \emph{via} fracture \cite{zhao2014multi,anderson2017fracture}.
In stiff materials, established theories allow one to predict failure of both brittle and ductile materials in arbitrary geometries, based on the results of a few standard mechanical tests \cite{griffith1921vi, dugdale1960yielding, anderson2017fracture}.
However, the same is not true of fracture in highly stretchable materials
\cite{knowles1977finite, arruda1993three, gent1996new, lakrout1999direct, baumberger2006solvent, seitz2009fracture, bouchbinder2009dynamic, livne2010near, bouklas2015effect, lee2019sideways, zhang2021relationship, li2021effects, yin2021peel, persson2021simple, wang2022hidden}.
The challenge, in this case, originates from the large deformation near the tip of a crack, which leads to nonlinear stress and strain fields as well as complex failure and dissipation processes near the crack tip \citep{creton2016fracture, long2015crack, long2021fracture}.
An emerging consensus is that soft fracture is controlled by two material length-scales: the size of the crack-tip failure zone where microscopic damage occurs, and the size of the non-linear zone, where non-linear elasticity dominates the deformation \cite{bavzant1997scaling, shull2004deformation, creton2016fracture, chen2017flaw, long2021fracture}.
The relative size of these to each other, and in comparison to sample dimensions ({\it e.g.} thickness and crack-length) dictates the form of the fracture process.

In this work, we reveal that this picture misses some essential features of fracture in strain-stiffening elastomers.
Specifically, we show that the fracture process involves the interaction of three independent cracks: one inner crack, and two edge cracks that grow inward from the faces of the sample.
The interaction of these three cracks increases the strength of thick samples.
This thickness-dependent behavior is governed by a dimensionless parameter reflecting the strain-stiffening characteristic of an elastomer.

\begin{figure}
\includegraphics[width=0.48\textwidth]{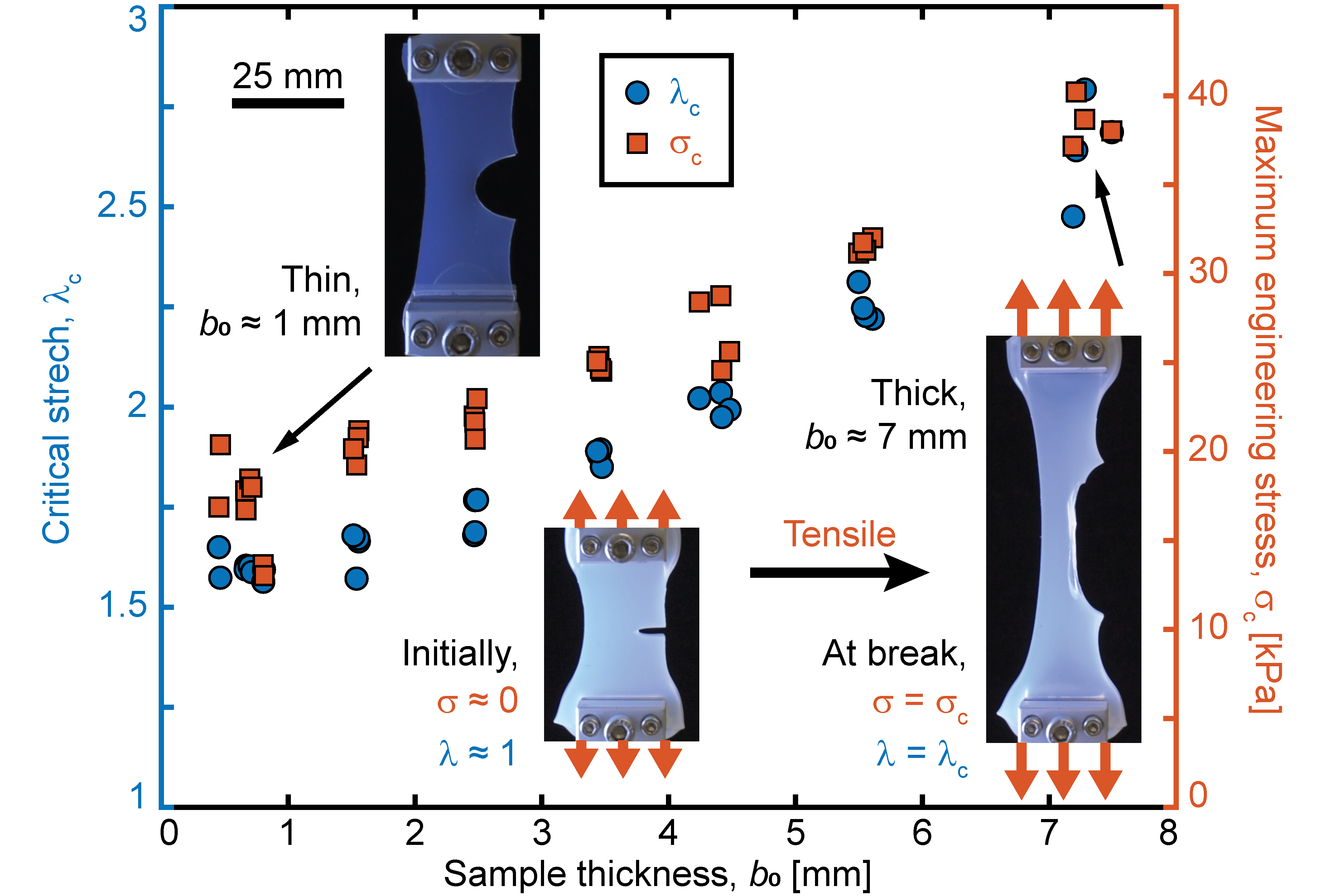}%
\caption{
The critical stretch, $\lambda_\mathrm{c}$ (blue circles), and maximum engineering stress, $\sigma_\mathrm{c}$ (red squares), of samples measured in single-edge-notch tension tests, as a function of the sample thickness $b_0$.
Stress, $\sigma$, is calculated by dividing the applied force by the initial cross-section of un-notched samples, while $\lambda$ is calculated by dividing the current sample length by its initial length.
Test samples are $30.5~\mathrm{mm} \times 25~\mathrm{mm} \times b_0$ (length $\times$ width $\times$ thickness).
The initial crack length is $10~\mathrm{mm}$, cut with a razor blade.
}
\label{fig:SEC}
\end{figure}

\begin{figure*}[]
\includegraphics[width=1\textwidth]{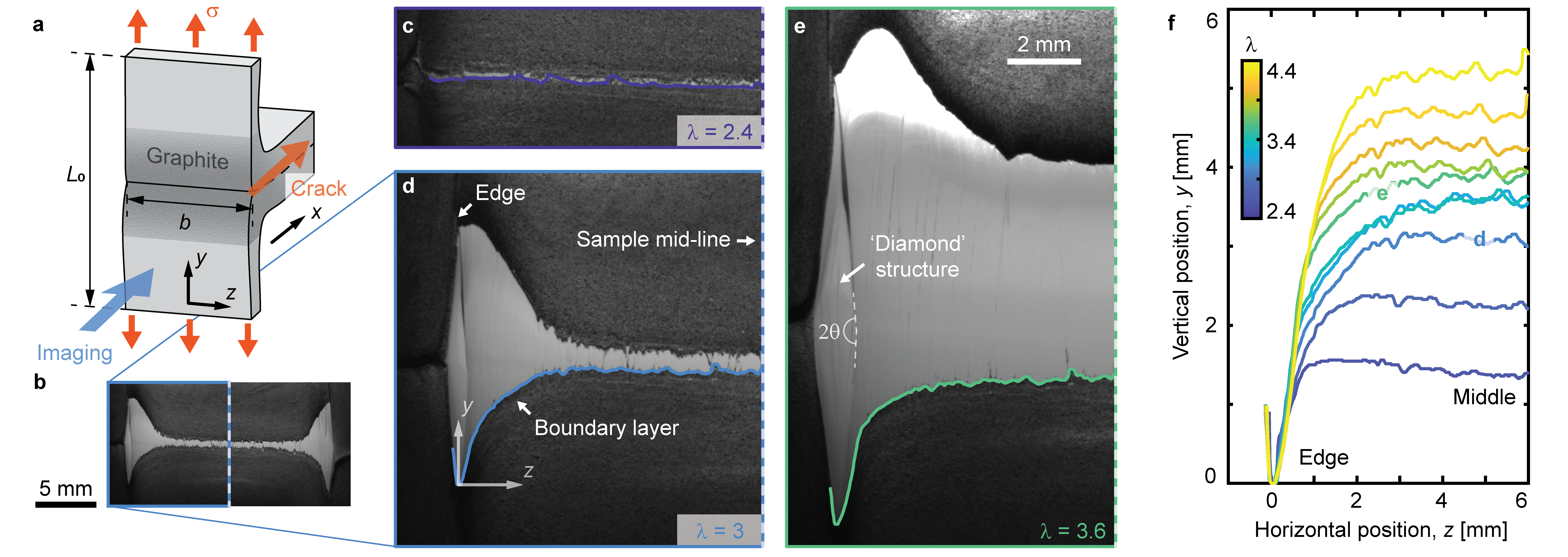}%
\caption{
Observing new fracture surface in T-peel tests.
({\bf a}) A schematic of the T-peel tests. ({\bf b}) A typical image of new fracture surface (dark area is graphite powder, bright area is new fracture surface).
All tests are symmetric about their mid-line (white, dashed line) to good approximation. Here, $b_0 = 25~\mathrm{mm}$.
({\bf c}-{\bf e}) Images of the left half of the same crack at increasing stretch: $\lambda = 2.4$, $3$, and $3.6$.
The continuous, colored curves denote the bottom boundaries of the new opening area.
({\bf f}) The curves of the bottom boundaries of the crack opening area at different stretches collapse onto a single curve near the sample edge, reminiscent of a boundary layer in fluid mechanics.
}
\label{fig:peeling}
\end{figure*}

In a standard fracture test, we rely on results being thickness-independent to justify the measurement of material properties \cite{greensmith1963rupture, ducrot2014toughening, creton2016fracture, long2016fracture, li2021effects}.
However, the fracture process in soft elastomers can be surprisingly thickness-dependent \cite{lee2019sideways, yin2021peel} (Fig.~\ref{fig:SEC}).
We perform single-edge-notch tension tests on samples of a commercial, highly stretchable silicone elastomer, Ecoflex 00-30 (Smooth-On) (see Methods for details).
These have identical dimensions, other than thickness, $b_0$, which we vary between $0.5~\mathrm{mm}$ and $7.5~\mathrm{mm}$: a typical range for materials used in soft devices.
Figure~\ref{fig:SEC} shows the ultimate tensile strength (\emph{i.e.} maximum engineering stress), $\sigma_\mathrm{c}$, and the corresponding stretch, $\lambda_\mathrm{c}$, for samples tested to failure (typical $\sigma$--$\lambda$ curves are shown in SI Fig.~S1).
Interestingly, thicker samples are more than twice as strong as thinner samples.
This trend is opposite to that found in metals, where thinner samples are stronger \cite{astm2005standard, anderson2017fracture}.
Alongside changes in $\sigma_\mathrm{c}$ with $b_0$, we also see changes in crack morphology (Fig.~\ref{fig:SEC}, Movies 1 and 2) \cite{lee2019sideways}.
Thinner samples exhibit classical Mode-I fracture, while thicker samples have extremely blunted crack tips.
This blunting has been attributed to the phenomenon of `sideways cracking' \cite{lee2019sideways}, where the crack path curves to travel parallel to the direction of applied tension.

To understand why our results are so thickness-dependent, we image the fracture surface across the sample thickness.  This is facilitated by the use of a T-peel geometry \cite{rivlin1953rupture, haque2012super, creton2016fracture,long2016fracture, yin2021peel}, as shown in Fig.~\ref{fig:peeling}{\bf a},   which allows us to use a large range of thicknesses (widths), $b_0$, ranging from $1~\mathrm{mm}$ to $25~\mathrm{mm}$.
Samples are formed in a mold containing a thin, metal sheet, which separates the two legs of the samples.
The sheet is removed after curing, eliminating the need to cut an initial crack in the sample (see Methods \& Supplementary Information).
Samples are clamped at a distance of $15.9 \pm 0.3  ~\mathrm{mm}$ from the initial crack front on each of the legs.
Then, we initiate the test at a grip-to-grip separation of $L_0 = 30.5 \pm 0.1 ~\mathrm{mm}$.
We increase the grip-to-grip stretch, $\lambda$, at a rate of ${\dot \lambda} = 0.4~\mathrm{min}^{-1}$, which is slow enough that viscoelastic rate effects will be negligible (see rheology in Ref.~\citep{darby2022modulus}, and Supplementary Information).

To directly visualize the creation of new fracture surface, we coat the sample with a dense layer of graphite powder so that new crack surface can be easily identified by a lack of graphite.
Upon stretching, this area lies essentially flat in the $y$-$z$ plane, so we always record images along the $x$ direction (the coordinate axis is defined in Fig.~\ref{fig:peeling}{\bf a}).
During loading, newly generated interface is bright and clearly visible, as seen in the example in 
Fig.~\ref{fig:peeling}{\bf b} (Movie 3).
Importantly, this area is almost always symmetric about the mid-line of the sample (dashed line in Fig.~\ref{fig:peeling}{\bf b}).
Thus, for compactness, we only show half of our images when presenting the results below.

Our images show that cracks evolve in highly non-uniform manner across a sample's thickness.
For example, Figures~\ref{fig:peeling}{\bf c}-{\bf e} show how new surface appears for a sample with $b_0=25~\mathrm{mm}$.
The crack first opens uniformly across the width of the sample (see image at $\lambda = 2.4$).
However, as the stretch increases ($\lambda = 3$), the new surface is generated much faster at the outer edges of the crack.
Later on ($\lambda = 3.6$), the crack growth accelerates near the mid-plane of the sample.
Perhaps surprisingly, crack growth is almost never up-down symmetric (\emph{e.g.} Fig.~\ref{fig:peeling}{\bf e}).
This asymmetry reflects the presence of `sideways' cracking mentioned above \citep{lee2019sideways}, which deflects the crack front either toward the $+y$ or $-y$ directions.
Here, we always orient images so that cracks appear to deflect upward.

\begin{figure*}
\includegraphics[width=1\textwidth]{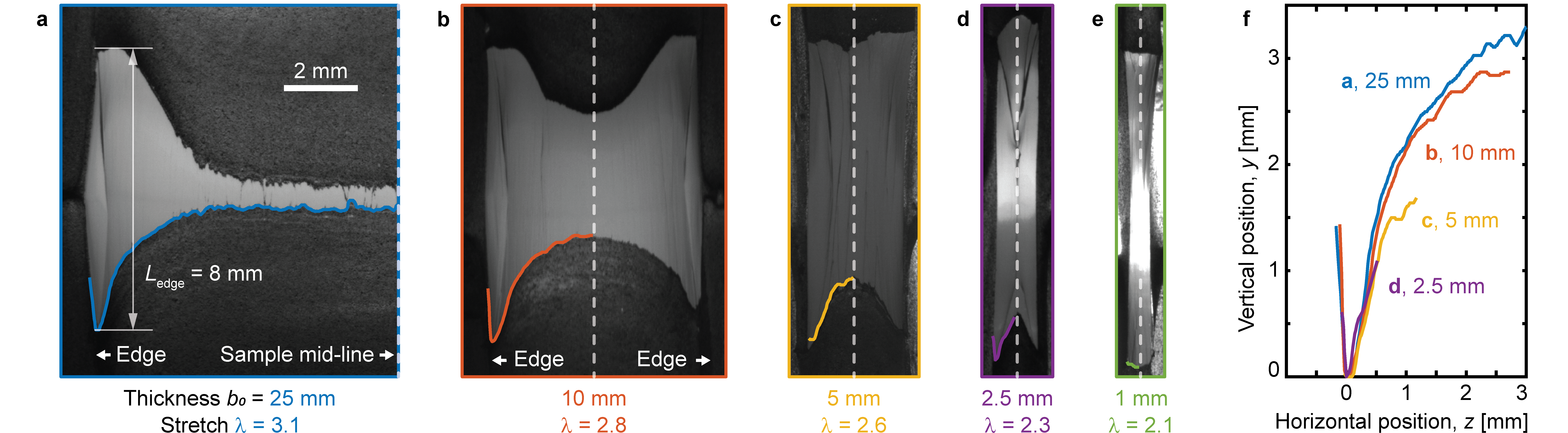}%
\caption{
The structure of boundary layers with different sample thicknesses. ({\bf a}-{\bf e}) Crack opening areas for samples with a range of different thicknesses, $b_0$, at the same edge opening length: $L_\mathrm{edge}=8~\mathrm{mm}$.
The continuous curves denote the bottom boundaries of the crack opening area, highlighting the `boundary layer' shape.
Dashed lines denote the vertical mid-lines of the samples.
({\bf f}) The bottom-boundary shapes from ({\bf a}-{\bf d}) collapse onto a single curve near the sample edge, supporting the idea of a well-defined boundary-layer structure that arises during fracture.
}
\label{fig:thickness}
\end{figure*}

\begin{figure*}
\includegraphics[width=1\textwidth]{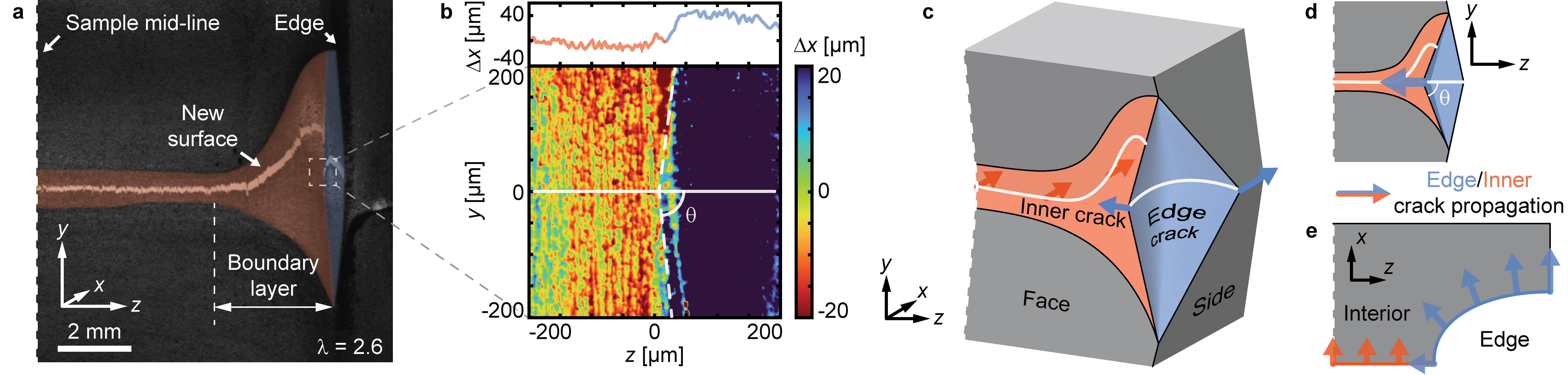}%
\caption{
The fracture surface is generated by three distinct cracks: an inner crack, and two edge cracks.
({\bf a}) Re-coating a fracture surface with graphite powder after it has started to grow allows us to see position of the crack tip (bright area).
This is discontinuous, showing that there are multiple cracks that grow into the sample simultaneously.
Inner cracks create the red-shaded opening area near the mid-plane of the sample.
Edge cracks create the diamond-shaped, blue-shaded area on the edges of the sample.
({\bf b}) Bottom: 3-D profile of the crack surface around the position where the edge-crack tip meets the inner crack. Top: a line-scan of the surface topology along the white, continuous line in the bottom image (passing through the edge-crack tip).
Dashed white lines indicate the boundary between the cracks, and $\theta$ is half the edge-crack tip-opening angle.
({\bf c}) 3-D schematic of the crack geometry. White curves show the crack tips, while arrows indicate local directions of crack propagation.
({\bf d},{\bf e}) The $y$-$z$ and $x$-$z$ projections of the schematic in ({\bf c}) respectively.
}
\label{fig:mechanism}
\end{figure*}

Cracks open uniformly near the mid-plane, while having an expanding edge structure with a fixed shape. 
Figure~\ref{fig:peeling}{\bf f} shows the shape of the bottom boundary of the developing crack at different stretches (colored curves in Figs.~\ref{fig:peeling}{\bf c}-{\bf e}).
We superpose these shapes by plotting them relative to the lowest point on the curves.
Interestingly, the shapes  collapse onto a single curve near the edge.
Outside of this `boundary layer', the profile  levels off, adopting a uniform opening at the center.
With further stretch, the boundary layer propagates deeper into the sample (see also SI Fig.~S4).
We borrow the term `boundary layer' from fluid mechanics, where flow profiles are uniform away from a surface, but are strongly affected by viscosity in a thin layer adjacent to the surface \cite{prandtl1905uber, batchelor2000introduction}.
In a similar fashion, the current boundary layer suggests a transition of the governing physics from the edge to the middle of the crack.

The boundary layer structure is independent not only of stretch but also of sample thickness.
Figures~\ref{fig:thickness}{\bf a}-{\bf e} (taken from Movies 3-7) show crack shapes in samples of different thicknesses, at the same edge opening, $L_\mathrm{edge} = 8~\mathrm{mm}$.
For all but the thinnest samples, the boundary-layer structures have the same shape near their tips, as demonstrated by plotting all the lower-boundary curves together (Fig.~\ref{fig:thickness}{\bf f}): a V-shape with a well-defined root angle.
For thinner samples, the boundary layers from the opposite edges of the crack overlap and the variation in crack opening from edge to mid-line becomes less pronounced.
The independence of the boundary-layer shape from sample thickness and applied stretch suggests that it is governed by a material property.
We test this hypothesis by visualizing the failure of a different material (Dragon Skin 30, Smooth-On, see SI Fig.~S5 and Movie 8).
There, we find a similar, stretch-independent boundary layer structure, but with a different profile.

The above boundary layer is actually the consequence of a 
structure that extends across the edge of the newly formed fracture surface, creating a diamond shape ({\it e.g.} Figs.~\ref{fig:peeling}{\bf d} and {\bf e}, and SI Fig.~S6).
The bottom vertex of the diamond coincides with the V-shaped tip of the boundary layer.
The width of the diamonds increases as the boundary layers grow in toward the sample mid-plane.
To characterize the diamond shapes, we measure the obtuse, internal angle, $2\theta$ nearest to the mid-plane (see Fig.~\ref{fig:peeling}{\bf e} and SI Fig.~S6).
Similar to the boundary-layer structure, $\theta$ is independent of sample thickness and applied stretch, but dependent on the material.
For Ecoflex, we find that $\theta = 86 \pm 1^\circ$, while for Dragon Skin, $\theta= 79 \pm 2^\circ$ (details in Methods).

\begin{figure*}[]
\includegraphics[width=1\textwidth]{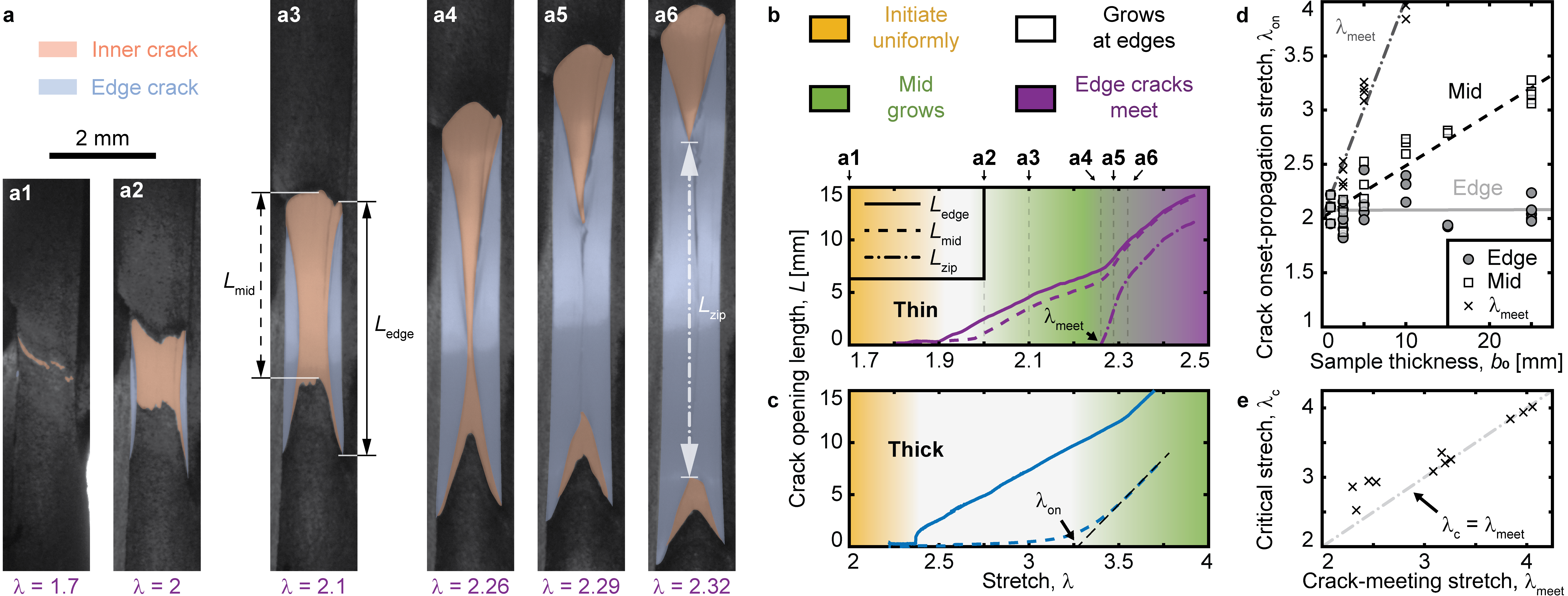}%
\caption{
Quantitative characterization of crack growth.
({\bf a}) Evolution of inner (red) and edge (blue) cracks for a sample with $b_0 = 2.5~\mathrm{mm}$.
({\bf b},{\bf c}) Crack opening lengths, $L_\mathrm{edge}$, $L_\mathrm{mid}$ \& $L_\mathrm{zip}$ (continuous, dashed, dash-dotted respectively) for thin ($b_0 = 2.5~\mathrm{mm}$) and thick ($b_0 = 25~\mathrm{mm}$, Fig.~\ref{fig:peeling}) samples.
Time-points shown in ({\bf a}) correspond to the points marked {\bf a1}-{\bf a6} in ({\bf b}).
Different colored regions indicate the four stages of crack growth: Yellow: a thin crack opens uniformly across the sample. White: edge cracks nucleate and start to grow. Green: the inner crack starts to grow. Purple: edge cracks meet at the sample mid-plane.
We extract effective stretches at the onset of crack propagation, $\lambda_\mathrm{on}$, with a linear fit to the initial stages of crack growth [{\it e.g.} dashed, black line in ({\bf c})].
({\bf d}) Stretches at the onset of crack propagation for edge cracks (circles) and the crack at the sample mid-plane (squares) as a function of the sample thickness $b_0$.
The two edge cracks meet at the sample mid-plane at $\lambda_\mathrm{meet}$ (crosses).
Lines show the best linear fits to the data sets.
({\bf e}) The stretch at maximum force, $\lambda_\mathrm{c}$, as a function of $\lambda_\mathrm{meet}$.
The dash-dotted line is $\lambda_\mathrm{c} = \lambda_\mathrm{meet}$.}
\label{fig:length}
\end{figure*}

The diamond structures are actually distinct edge cracks. 
We show this by first growing the fracture surface under an increasing stretch until $\lambda = 3$, and then re-applying a graphite powder coating after unloading the sample.
Upon further stretching, a new, bright area shows us the exact position of the crack tip (Fig.~\ref{fig:mechanism}{\bf a} and Movie 9).
Intriguingly, this new opening area does not connect the two sides of the sample in a single crack tip, but is discontinuous (Fig.~\ref{fig:mechanism}{\bf a}).
Across the majority of the thickness of the sample, there is a single opening strip, which curves upward as it approaches the sample sides, before abruptly stopping when reaching the diamond-shaped feature.
A new opening area then appears inside the diamond structure, cutting almost straight across its waist.
This observation suggests that there are actually three separate crack fronts across the sample thickness: one inner crack, and two edge cracks.
The inner crack generates new surface area across the middle of the sample (red-tinted area in Fig.~\ref{fig:mechanism}{\bf a}), while the edge cracks generate the diamond structures at the sample edges (blue-tinted area).

We visualize the transition from inner crack to edge crack by imaging the stretched surface with an optical profilometer.
Fig.~\ref{fig:mechanism}{\bf b} (bottom) shows the 3-D sample surface at the inner end of the edge crack (dashed box in Fig.~\ref{fig:mechanism}{\bf a}).
The blue area in this figure is the area created by the edge crack.
A height profile along the white line is shown at the top of the figure.
Both the inner-crack and edge-crack opening areas predominantly lie  flat in the $y$-$z$ plane.
However, the edge crack cuts deeper into the sample, as evidenced by the clear step in the surface profile at the interface between the two crack opening areas.

Altogether, our observations suggest an overall 3-D crack-tip morphology in Fig.~\ref{fig:mechanism}{\bf c}, with corresponding 2-D projections in Figs.~\ref{fig:mechanism}{\bf d} and {\bf e}.
Neglecting the `sideways' crack propagation in the $y$ direction, the inner crack propagates predominantly in the $x$ direction, while the edge crack propagates in the $x$ and $z$ directions.
This structure implies that the tip of the diamond corresponds to the tip of the edge crack
(Fig.~\ref{fig:mechanism}{\bf d}), with a shared opening angle, $2 \theta$.

Inner and edge cracks grow at different applied stretches.
This is clearly seen by comparing the edge and mid-plane opening lengths, $L_\mathrm{edge}$ and $L_\mathrm{mid}$, at different levels of stretch in different samples (\emph{e.g.} Fig.~\ref{fig:length}{\bf a}).
Just after uniform initiation across the whole sample (Fig.~\ref{fig:peeling}{\bf c}), $L_\mathrm{edge}$ and $L_\mathrm{mid}$ maintain similar small values (yellow regions of  Fig.~\ref{fig:length}{\bf b}-{\bf c}).
Then, the edge cracks start to propagate rapidly, while the inner crack stays stationary (white region).
After a lag, the inner crack starts to propagate (green region).
This lag is larger in thicker samples.
To quantify this, we measured the sample stretch at the onset of crack propagation, $\lambda_\mathrm{on}$, determined by extrapolating the rapid linear growth regime back to zero length (see Fig.~\ref{fig:length}{\bf c}).
The onsets of edge- and inner-crack propagation are shown for a range of sample thicknesses in  Fig.~\ref{fig:length}{\bf d}.

Interestingly, the onset of edge-crack propagation appears to be independent of thickness, while inner cracks require larger stretches to propagate in thicker samples.
Edge cracks start to propagate at a constant stretch of about 2, independent of $b_0$.
This suggests that edge cracks are strictly associated with the surface.
By contrast, the onset of inner-crack propagation  increases linearly with thickness.
This suggests that the onset of inner-crack propagation is affected by the inward-propagation of edge-cracks.

The samples fail when the edge cracks meet, at a stretch $\lambda_\mathrm{meet}$.
At this point, under controlled-displacement loading, the two edge cracks `zip' together.
The length of boundary between the two edge cracks, $L_\mathrm{zip}$ (\emph{e.g.} Fig.~\ref{fig:length}{\bf a}6), is plotted as a function of stretch in Fig.~\ref{fig:length}{\bf b}.
The zipping of edge cracks accelerates the opening of new fracture surface, indicated by the kinks in the curves of $L_\mathrm{mid}$ and $L_\mathrm{edge}$ at this point.
Crucially, the force applied to the sample either abruptly plateaus, or drops off when edge cracks meet, as we see by comparing the stretch at maximum force, $\lambda_\mathrm{c}$ with $ \lambda_\mathrm{meet}$ in Fig.~\ref{fig:length}{\bf e} (see also SI Fig.~S7).
Thus, edge-crack meeting defines the sample's load-bearing capacity.

These results suggest an empirical criterion for failure, based on the meeting of edge cracks.
Given the consistent diamond-shape of the edge cracks, the distance they move inward is proportional to edge-crack opening. 
Ignoring corrections due to the Poisson effect,  edge cracks reach the mid-plane when $L_\mathrm{edge}\approx b_0\tan\theta$, about $10 b_0$ for Ecoflex.
Conveniently, data in Figs.~\ref{fig:length}{\bf b}-{\bf d} show that edge-crack opening is linear with stretch above the onset of propagation.
Thus, $L_\mathrm{edge}/L_0 \approx c (\lambda-\lambda_\mathrm{on}^\mathrm{edge})$, where $c$ is a constant of $O(1)$ that depends weakly on the material and thickness (SI Fig.~S8).
Combining these, we find that $\lambda_c\approx \lambda_\mathrm{on}^{\mathrm{edge}} + b_0 \tan \theta /(c L_0)$.
This failure criterion increases linearly with thickness, consistent with our observations in single-edge-notched tension (Fig.~\ref{fig:SEC}) and T-peel tests (Fig.~\ref{fig:length}{\bf d} and SI Fig.~S7).

The form of this failure criterion highlights the importance of the geometric ratio, $L_0/b_0$.
When $L_0/b_0\ll \tan \theta$ (the limit of thick samples), edge cracks need to propagate significantly before meeting in the mid-plane.
This leads to thickness-dependent, higher critical stretches.
In contrast, when $L_0/b_0\gg \tan \theta$ (the limit of thin samples), edge cracks meet as soon as they form, and failure occurs at $\lambda_{\mathrm{on}}^{\mathrm{edge}}$, independent of thickness.
Extending this argument, we expect that fracture tests will be thickness-dependent, unless all of the ratios of lateral sample dimensions to $b_0$ are much greater than $\tan \theta$.
Applying these ideas to the single-edge-notch tension tests in Fig. \ref{fig:SEC}, we identify the initial crack length ($10~\mathrm{mm}$) as the smallest lateral dimension, as it is smaller than the ligament length, and grip-to-grip spacing.
Thus, we require $b_0\ll 10~\mathrm{mm}/\mathrm{tan}(86^\circ) \approx 0.7~\mathrm{mm}$ for thickness-independent behavior, in reasonable agreement with our observations.  

This last argument is based on two experimental observations: 1) edge cracks propagate at a lower stretch than the inner crack, and 2) edge cracks maintain a consistent diamond shape.
Both of these facts can be rationalized by reinterpreting some previous results of fracture mechanics for nonlinear-elastic materials.

Cracks propagate more easily at the edge because they concentrate stress more severely there.
This follows from the fact that the analytical, crack-tip stress singularity in the deformed configuration is stronger for plane stress (\emph{i.e.} edge cracks) than for plane strain (\emph{i.e.} inner cracks) for a range of nonlinear-elastic materials (see Ref.~\citep{long2015crack} and Supplementary Section C).
Intriguingly, the opposite behavior is seen in metals, where cracks grow first in the middle \cite{mises1913mechanik, kudari20173d, anderson2017fracture}.

The robust diamond shape of the edge cracks follows from the material's nonlinear-elastic behavior: in sufficiently strain-stiffening materials, crack tips take wedge-like shapes with a constant opening angle, $2\theta$.
Tensile tests reveal that the two silicones used in this study are strain stiffening, and well-fit by an exponential  model with strain-energy density function $W = \mu J_\mathrm{m}\left[e^{(I_1-3)/J_\mathrm{m}}-1 \right]/2$ (SI Fig.~S9).
Here, $\mu$ is the shear modulus, $I_1$ is the trace of the right Cauchy-Green deformation tensor \cite{seitz2009fracture, cristiano2010experimental}, and $J_\mathrm{m}$ is a dimensionless material parameter.
This reduces to the familiar Neo-Hookean energy density in the limit of large $J_\mathrm{m}$. 
For Ecoflex and Dragon Skin, $J_\mathrm{m}= 36.5 \pm 0.4$ and $14 \pm 2$ respectively (SI Fig.~S9).
These values control the crack-tip angle, $\theta$, as theory predicts that $\tan \theta = \alpha J_\mathrm{m}^{3/4}$, where $\alpha$ is an $O(1)$ constant (see Refs.~\citep{long2011finite, long2015crack} and Supplementary Information).
Indeed, we find consistent values for our materials: for Ecoflex, $\alpha^\mathrm{Eco} = 0.9 \pm 0.2$, and for Dragon Skin, $\alpha^\mathrm{Dra} = 0.7 \pm 0.3$.
In the absence of strain stiffening (Vytaflex 40 polyurethane, Smooth-On \cite{moser2022hydroelastomers}), we observe no boundary layer or diamond structure (SI Fig.~S10).
Thus, strain stiffening appears to control edge crack behavior.

In conclusion, we have shown that the results of fracture tests on strain-stiffening elastomers can be thickness dependent, even for `thin' samples such as might be used in standard tests.
Reversing the familiar thickness dependence of metals, thicker strain-stiffening elastomers appear to be stronger than thinner ones \cite{yin2021peel}. 
The underlying cause is the strain-stiffening of the material, which leads to a fracture surface comprised of three independent cracks:
an inner crack initiates first, but propagates slowly; 
two diamond-shaped edge cracks have a delayed initiation, but propagate more easily.
When the edge cracks meet at the sample mid-plane, the sample fails.

The emerging view of soft-solid fracture has revolved  around two material length scales -- the sizes of the failure  and nonlinear-elastic zones \cite{long2021fracture}.
For the Ecoflex elastomer studied here, they are $O(0.1)~\mathrm{mm}$ and $O(1)~\mathrm{mm}$, respectively (see Supplementary Information).
Our work implies that these two length scales are not sufficient to describe the fracture process.
Edge cracks are an essential and unexpected feature of the nonlinear-elastic zone, arising from a material's strain-stiffening response.
Thus, a proper description of the fracture response must go beyond the familiar material length scales, and incorporate a dimensionless strain-stiffening parameter, such as $J_m$.

Generalizing our results, we expect that sample thicknesses must be much smaller than the smallest lateral sample dimension divided by $J_\mathrm{m}^{3/4}$ in order to achieve fracture tests with meaningful, thickness-independent results.
Since strain-stiffening materials only get stronger as their thickness increases, fracture results in the thin-sample limit can serve as a convenient lower bound of fracture strength for device design.
Challenges for future work include validation of our results over a wider class of strain-stiffening materials, and the impact of related phenomena, like strain-induced crystallization \cite{trabelsi2002stress, lee2019sideways}.  
While $J_\mathrm{m}$ reveals how long a material can survive after the onset of propagation of an edge crack, a complete understanding of the failure of these materials requires  elucidation of the factors that drive edge-crack initiation and their onset of propagation.


\section*{Methods}


We create Ecoflex 00-30 (Smooth-On) and Dragon Skin 30 (Smooth-On) samples by successively mixing together the two components (Parts A and B) in a $1:1$ ratio, degassing, and curing at $40^\circ\mathrm{C}$ for 24 hours.
Before mixing the components together, these are centrifuged at $10^4 \times$ gravity for 4 hours to remove any large clumps of suspended silica particles.
This minimizes sample heterogeneity.
Samples are used within 12 hours after finishing curing.
Vytaflex 40 (Smooth-On) samples are simply mixed in a $1:1$ ratio, and allowed to cure at room temperature for 24 hours.


Single-edge-notch tension specimens, are created by curing slabs of Ecoflex 00-30 in Petri dishes, and cutting out the desired shapes (see SI Fig.~S1{\bf a}).
These are tested with a tensile-testing machine (Stable Micro Systems, TA.XTPlus, $5~\mathrm{kg}$ load-cell).
The initial grip-to-grip distance is $L_0 = 30.5 \pm 0.1 ~\mathrm{mm}$, and tests are performed at a constant stretch rate of ${\dot \lambda} = 0.4~\mathrm{min}^{-1}$ ({\it e.g.} SI Fig.~S1{\bf b}).

T-peel samples are cured in a laser-cut acrylic mold containing a 50\,$\mu$m-thick, molybdenum sheet (see SI Fig.~S2{\bf a}) that separates the two sample legs.
The use of this sheet avoids the need for cutting the samples to form an initial crack.
The sheet is removed after curing, and then the legs of the sample are clamped in the tensile-testing machine for testing.
For thin samples ($b_0 \leq 2.5~\mathrm{mm}$), we remove excess weight from the back of the sample with a razor blade (see the dashed line in SI Fig.~S2{\bf a}).

We image new fracture surface by coating graphite powder ($5~\mathrm{\mu m}$, Sigma-Aldrich) on the sample surfaces with a brush.
These surfaces are imaged with a camera (Thorlabs, SC1280G12M) with telecentric lens (Seiwa Optical, FXL-0305-VT-165, $0.3\times$).
The sample is imaged in transmission with an LED panel placed behind the sample.
The 3D profile of the crack surface in Fig.~\ref{fig:mechanism}{\bf b} is measured with a 3D optical profilometer (S-neox, Sensoscan, $20 \times$ objective).
Here, no graphite powder is applied, and the sample is directly stretched to $\lambda = 2.6$.

Edge-crack opening angles, $2 \theta$, are measured in T-peel experiments with 21 measurements from four experiments with $b_0 = 10~\mathrm{mm}$ or $25~\mathrm{mm}$  (Ecoflex) and with 16 measurements from four experiments with $b_0 = 20~\mathrm{mm}$ (Dragon Skin).


We characterize strain-stiffening properties with multiple uniaxial tensile tests on rectangular samples.
The samples are approximately $12.5~\mathrm{mm}$ wide and $1.5~\mathrm{mm}$ thick (cut by a razor blade), and clamped in the tensile-testing machine with an initial grip-to-grip distance of $L_0 = 15.5~\mathrm{mm}$.
Samples are stretched at a constant stretch rate ${\dot \lambda} = 0.4~\mathrm{min}^{-1}$, and we calculate the strain-energy density $W(\lambda)=\int_1^\lambda \sigma d\lambda$, where $\sigma$ is the engineering stress, and $\lambda$ is stretch.
This is then fitted to the exponential model given in the article by setting $I_1=\lambda^2+2/\lambda$ (see examples in SI Fig.~S9).

\section*{Data Availability}
The datasets and the Matlab codes to support this study will be available online and in open source.

\section*{Acknowledgement}
The authors thank Nikolaos Bouklas and Chung-Yuen Hui at Cornell University for inspiring discussions.
We thank Lucio Isa and Tobias Gm\"ur at ETH Z\"urich for helping the use of profilometer.
R.L. acknowledges support from the United States National Science Foundation (NSF CMMI-1752449).

\section*{Author Contributions}
R.W.S. initiated this work.
N.X. performed the experiments and processed the data.
N.X., R.L., E.R.D., and R.W.S. discussed the results and wrote the paper.

\section*{Conflicts of interest}
The authors declare no competing interests.

\end{document}